\pdfoutput=1
\documentclass[10pt,conference,letterpaper]{IEEEtran}
\IEEEoverridecommandlockouts
\usepackage{mathrsfs}
\usepackage{pifont}
\usepackage{bbding}
\usepackage{multirow}
\usepackage{url}
\usepackage{amssymb}
\usepackage{stfloats}
\usepackage{amssymb}
\usepackage{makecell}
\usepackage{booktabs}
\usepackage{graphicx} %use graph format
\usepackage{epstopdf}
\usepackage{threeparttable}

\usepackage{graphicx}
\usepackage{amsthm}
\usepackage{balance}
\usepackage{amssymb, amsfonts, epsfig, latexsym, times}
\usepackage{bm}
\usepackage{color}
\usepackage{cite}
\usepackage{algorithm, algorithmic}
\usepackage{graphicx}
\usepackage{caption}
\usepackage{float}
\usepackage{subcaption}
\usepackage[tbtags]{amsmath}
\usepackage[justification=raggedright,singlelinecheck=false,font=footnotesize]{caption}
\usepackage[subrefformat=parens,labelformat=simple,justification=centerlast]{subcaption}
\IEEEoverridecommandlockouts
\def\BibTeX{{\rm B\kern-.05em{\sc i\kern-.025em b}\kern-.08em
    T\kern-.1667em\lower.7ex\hbox{E}\kern-.125emX}}
\usepackage{empheq}
\usepackage[letterpaper,right=0.635in,left=0.635in,top=0.8in,bottom=1.1in]{geometry}

\usepackage{acro}
\DeclareAcronym{isac}{
  short = ISAC,
  long  = Integrated Sensing and Communication
}

\DeclareAcronym{dfrc}{
  short = DFRC,
  long  = Dual Functional Radar and Communication
}
\DeclareAcronym{ra}{
  short = RA,
  long  = Reconfigurable Antenna
}
\DeclareAcronym{rf}{
  short = RF,
  long  = Radio Frequency
}

\DeclareAcronym{em}{
  short = EM,
  long  = Electromagnetic
}

\DeclareAcronym{fd}{
  short = FD,
  long  = Fully Digital
}
\DeclareAcronym{mu}{
  short = MU,
  long  = Multi-user
}

\DeclareAcronym{mimo}{
  short = MIMO,
  long  = Multiple-input Multiple-output
}

\DeclareAcronym{miso}{
  short = MISO,
  long  = Multiple-input Single-output
}
\DeclareAcronym{cus}{
  short = CUs,
  long  = Communication Users
}
\DeclareAcronym{cu}{
  short = CU,
  long  = Communication User,
  first-style = short
}
\DeclareAcronym{hbf}{
  short = HBF,
  long  = Hybrid Beamforming
}
\DeclareAcronym{dof}{
  short = DoF,
  long  = Degree of Freedom
}
\DeclareAcronym{dofs}{
  short = DoFs,
  long  = Degrees of Freedom
}
\DeclareAcronym{era}{
  short = ERA,
  long  = Electromagnetically Reconfigurable Antenna
}
\DeclareAcronym{oa}{
  short = OA,
  long  = Omnidirectional Antenna
}

\DeclareAcronym{ba}{
  short = BA,
  long  = Broadside Antenna
}

\DeclareAcronym{cgv}{
  short = CGV,
  long  = Complex Channel Gain Vector
}
\DeclareAcronym{agv}{
  short = AGV,
  long  = Antenna Gain Vector
}
\DeclareAcronym{arv}{
  short = ARV,
  long  = Antenna Response Vector
}
\DeclareAcronym{ue}{
  short = UE,
  long  = User Equipment
}

\DeclareAcronym{ula}{
  short = ULA,
  long  = Uniform Linear Array
}

\DeclareAcronym{upa}{
  short = UPA,
  long  = Uniform Planar Array
}

\DeclareAcronym{los}{
  short = LoS,
  long  = Line-of-Sight
}
\DeclareAcronym{aod}{
  short = AoD,
  long  = Angle-of-Departure
}

\DeclareAcronym{awgn}{
  short = AWGN,
  long  = Additive White Gaussian Noise
}

\DeclareAcronym{sinr}{
  short = SINR,
  long  = Signal-to-Interference-plus-Noise Ratio
}
\DeclareAcronym{scnr}{
  short = SCNR,
  long  = Signal-to-Clutter-plus-Noise Ratio
}
\DeclareAcronym{mse}{
  short = MSE,
  long  = Mean Squared Error
}
\DeclareAcronym{bcd}{
  short = BCD,
  long  = Block Coordinate Descent
}

\DeclareAcronym{ao}{
  short = AO,
  long  = Alternating Optimization
}

\DeclareAcronym{fp}{
  short = FP,
  long  = Fractional Programming
}

\DeclareAcronym{mo}{
  short = MO,
  long  = Manifold Optimization
}

\DeclareAcronym{sip}{
  short = SIP,
  long  = Semi-Infinite Programming
}

\DeclareAcronym{rmi}{
  short = RMI,
  long  = Radar Mutual Information
}

\DeclareAcronym{sandc}{
  short = S\&C,
  long  = Sensing and Communication
}

\DeclareAcronym{itu}{
  short = ITU,
  long  = International Telecommunication Union
}

\begin{document}
\captionsetup[figure]{labelsep=period,singlelinecheck=off,font=small}
\captionsetup[table]{
	labelsep = newline,%period:点号, space:空格, : , newline:换行
	singlelinecheck=true,%居中
	font=small,%字体
}
\title{Integrated Sensing and Communication with Tri-Hybrid Beamforming Across Electromagnetically Reconfigurable Antennas}
\author{\IEEEauthorblockN{Jiangong Chen\IEEEauthorrefmark{1}, Xia Lei \IEEEauthorrefmark{1}, Yuchen Zhang\IEEEauthorrefmark{2}, Kaitao Meng\IEEEauthorrefmark{3}, and Christos Masouros\IEEEauthorrefmark{4}}
	\IEEEauthorblockA{
		\IEEEauthorrefmark{1}National Key Lab. of Wireless Communications, University of Electronic Science and Technology of China, China\\}
	\IEEEauthorblockA{
		\IEEEauthorrefmark{2} CEMSE, King Abdullah University of Science and Technology, Thuwal, Saudi Arabia\\}
	\IEEEauthorblockA{\IEEEauthorrefmark{3} Department of Electrical and Electronic Engineering, University of Manchester, Manchester, UK \\}
    \IEEEauthorblockA{\IEEEauthorrefmark{4}Department of Electronic and
Electrical Engineering, University College London, London, UK\\}
	\IEEEauthorblockA{
		Email: \{jg\_chen@std.uestc.edu.cn\}}
}

\maketitle
\begin{abstract}
Beamforming with a sufficient number of antennas is one of the most significant technologies for both \ac{mu} \ac{mimo} communication and \ac{mimo} radar sensing in \ac{isac} systems. However, its performance suffers from limited \ac{dofs} in conventional hybrid beamforming systems. To overcome this, we propose an \ac{era}-aided \ac{isac} system, where transmit \ac{era}s dynamically adjust their radiation patterns to enhance system DoFs and improve overall performance. Specifically, we design a tri-hybrid beamforming optimization framework combining digital, analog, and \ac{em} beamforming to jointly maximize communication rate and sensing \ac{scnr}. Furthermore, an integrated \ac{fp} and \ac{mo} approach is developed to transform the problem into tractable subproblems with closed-form updates. Simulation results verify that the proposed ERA-ISAC system achieves almost 10 dB \ac{sandc} performance gain compared to its conventional hybrid beamforming counterparts with \ac{oa}.

% To solve this problem, we develop an optimization framework integrating extended \ac{fp} and \ac{mo}. Unlike classic \ac{fp} approaches, we decouple the \ac{scnr} term into tractable fractional forms, applying \ac{fp} transformations to linearize the objective and convert the original problem into a block-wise convex structure with closed-form solutions per subproblem. Furthermore, to address the non-convex constant-modulus and norm constraints inherent in analog and EM beamforming, we rigorously derive the corresponding Riemannian gradients and perform optimization on the appropriate manifolds. Simulation results verify that the proposed ERA-ISAC system achieves better \ac{sandc} performance compared to its conventional hybrid beamforming counterparts with \ac{oa}.

\end{abstract}
\acresetall
\IEEEpeerreviewmaketitle
\section{Introduction}

% Owing to its powerful spatial diversity and multiplexing capabilities, \ac{mimo} technology has witnessed rapid development in recent years, driving a wide range of emerging applications such as spatial multiplexing \cite{sdma}, physical layer security \cite{dm,an}, unmanned aerial vehicle communications \cite{uav}, and so on. \ac{isac}, as one of these applications, has been listed as one of the six defining features of 6G by \ac{itu} in International Mobile Telecommunications (IMT)-2030 framework. Over the past decades, a large number of works have focused on the co-existence or even joint transmission of \ac{mu}-\ac{mimo} precoding/beamforming and \ac{mimo} radar sensing \cite{fanliudfrc}. In these systems, large-scale antenna arrays are capable of generating highly directive beams, which are beneficial for enhancing both communication and sensing performance. 

Owing to the intrinsic spatial diversity and multiplexing gains, \ac{mimo} has been regarded as a fundamental enabler for \ac{isac}, allowing large-scale arrays to synthesize highly directive beams that jointly enhance both communication and sensing performance. This synergy has motivated extensive studies on the co-existence and joint design of \ac{mu}-\ac{mimo} precoding/beamforming with \ac{mimo} radar sensing \cite{fanliudfrc}. However, a large number of antennas and \ac{rf} chains also pose significant complexity in terms of hardware overhead and power consumption \cite{MassiveMIMO}. To alleviate implementation complexity, the \ac{hbf} architecture has been extensively studied, where \ac{dofs} are exchanged for a reduction in hardware cost \cite{HBF}. How to further enhance the \ac{dofs} and performance of \ac{mimo} systems while maintaining the low-RF-chain characteristics of \ac{hbf} systems remains an urgent challenge. This issue becomes particularly critical when the number of antennas is small, as the available \ac{dofs} in both digital and analog beamformers decrease sharply in such scenarios.

Against this background, fluid/movable antennas \cite{FAHBF1, FAHBF}, which flexibly tune antennas' positions to reconfigure the wireless channel towards a more favorable condition, have recently been proposed and demonstrated remarkable performance gains. These two antenna technologies suffer from two primary drawbacks. Firstly, their reliance on hardware mechanisms like moving parts or liquid metal results in slow speed, high cost, and reliability concerns. Secondly, its use of positional reconfigurability to alter the effective array pattern, rather than enabling independent control over individual antenna patterns, fundamentally limits its beamforming \ac{dofs}. Beyond position reconfiguration, broader forms of reconfigurability have been realized through recently emerging \ac{ra}, enabling dynamic adjustment of each antenna element’s electromagnetic properties at the transceiver, such as radiation pattern, polarization, operational frequency, and even bandwidth. Correspondingly, the hybrid-\ac{mimo} has been extended to the tri-hybrid-\ac{mimo} architecture \cite{trimimo}. At present, research on \ac{ra} has primarily focused on antenna design and communication systems. For example, in \cite{RA1,RA2,RA3,RA4}, the authors exploited the radiation pattern reconfiguration capability of \ac{ra} for precoding and channel estimation, while \cite{RAPol} employed the \ac{ra} for polarforming. However, few studies have considered exploiting the additional \ac{dofs} introduced by \ac{ra} to strike a superior trade-off performance in \ac{isac} systems. Therefore, in this paper, we follow the \ac{era} model in \cite{RA1,RA2,RA3} based on spherical harmonics and consider its application in \ac{isac} systems. This paper makes several key contributions. First, it represents one of the earliest efforts to introduce \ac{era}s to \ac{isac} systems, underscoring their potential to enhance \ac{sandc} performance. To exploit the additional \ac{dofs} offered by \ac{era}s, we formulate a tri-hybrid beamforming (including baseband digital, phase-shifter analog, and \ac{era} \ac{em} beamforming) optimization problem to maximize the weighted sum of the \ac{mu} sum rate and sensing \ac{scnr}. To efficiently solve this highly nonconvex problem, we develop an \ac{ao} framework integrated with an extended \ac{fp} technique that linearizes the nonlinear sensing \ac{scnr} term by decomposing it into tractable fractional forms. This transformation converts the original objective into a block-wise convex structure, enabling closed-form updates for each variable block. Furthermore, \ac{mo} is employed to handle the constant-modulus and norm constraints of the analog and EM beamforming.

\section{System Model and Problem Formulation}
\subsection{System and Channel Model}
\subsubsection{System Model}
\begin{figure}
    \centering
	\includegraphics[scale=0.7]{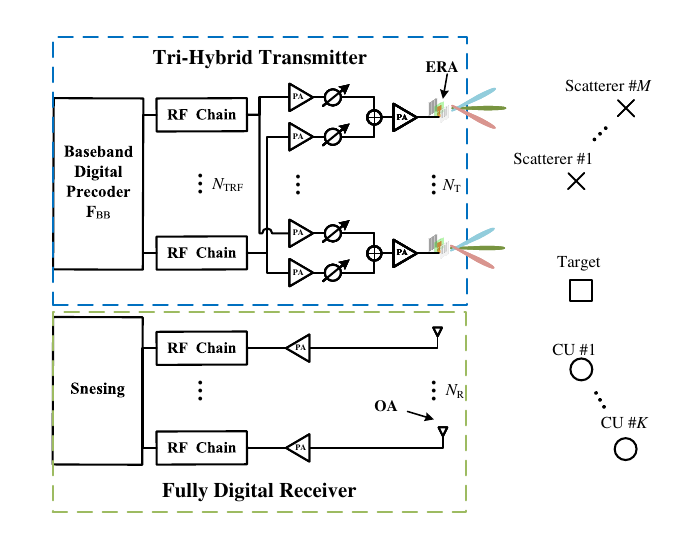}
	 \caption{System model of the proposed ERA-ISAC.}
	\label{SystemModel}
\end{figure}
As depicted in Fig. \ref{SystemModel}, we consider a downlink \ac{mu} \ac{miso} \ac{isac} system including a transmitter equipped with an $N_{\rm T}$-element \ac{upa}, $N_{\rm R}$-element \ac{upa} radar receiver, $K$ single-antenna \ac{cus} located at $(\theta_{k,l},\phi_{k,l})$, and one point sensing target located at $(\theta_{\rm t},\phi_{\rm t})$. The transmitter is of a fully-connected \ac{hbf} architecture with $N_{\rm TRF}$ RF chains and $N_{\rm T}$ \ac{era}s. This hardware architecture provides \ac{dofs} for digital, analog, and \ac{em} precoding/beamforming, which is named tri-hybrid beamforming. Since the detection mission is not the primary focus of this work, a fully digital architecture and \ac{oa}s are assumed for the radar receiver.
\subsubsection{Communication Channel Model}
According to \cite{RA1}, under far-field conditions, a general model applicable to arbitrary antenna types and  array shapes can be expressed as
\begin{equation}
\mathbf{h}_k = \sqrt{ \frac{N_{\rm T}}{L_k}}\sum_{l = 1}^{L_k} {\alpha}_{k,l} \cdot \mathbf{g}_{k,l} \odot \mathbf{a}_{k,l},
\end{equation}
where subscripts $k$ and $l$ are used to index \ac{cu} and path, $L_k$ denotes the path number associated with the $k$th \ac*{cu}, $\alpha_{k,l}$ denotes the corresponding complex channel gain \cite{RCS}, and $\odot$ represents the Hadamard product operation. The \ac{agv} $\mathbf{g}_{k,l}$ and \ac{arv} $\mathbf{a}_{k,l}$ all corresponding to the $l$th path of the $k$th \ac{cu} are defined as
\begin{itemize}
    \item AGV describes the radiation pattern of each antenna as
    \begin{equation}
        \mathbf{g}_{k,l} = [G_{k,l}^{(1)},\,G_{k,l}^{(2)},\,\dots,G_{k,l}^{(N_{\rm T})}]^{\rm T} \in \mathbb{C}^{N_{\rm T}}.
    \end{equation}
    For the transmitting channel vector, the elements of \ac{agv} are different since we employ \ac{era}s, but they are the same for the receiving channel vector due to the use of \ac{oa}s.
    \item ARV describes the phase differences between different antennas in an antenna array. Different array geometries exhibit distinct ARVs. A \ac{upa} located in the XoY plane is employed in this work for both transmitting and receiving arrays, and the \ac{arv} is given by
    \begin{equation}
        \begin{aligned}
                    \mathbf{a}^N_{k,l}\left(\theta_{k,l},\phi_{k,l}\right) = e^{-j 2 \pi / \lambda [0:N^x-1] d_x \sin \theta_{k,l} \cos \phi_{k,l} } \otimes \\
        e^{-j 2 \pi/\lambda [0:N^y-1] d_y \sin \theta_{k,l} \sin \phi_{k,l} },
        \end{aligned}
    \end{equation}
    where $N^x$ and $N^y$ are the antenna number along $x$ and $y$ axis, $d_x$ and $d_y$ are the corresponding antenna element spacing, $\otimes$ denotes the Kronecker product operation.
\end{itemize}

Focusing on the specific expression of the \ac{agv}, according to \cite{RA1}, the radiation pattern of the $n$th \ac{era} can be decomposed into an infinite summation of spherical harmonics as follows
\begin{equation}\label{NoAppG}
    G^{(n)}_{k,l} = G^{(n)}\left(\theta_k,\phi_{k,l}\right) = \sum_{u=0}^{+\infty} \sum_{q = -u}^{u} c_{uq}^{(n)}Y_u^{q}\left(\theta_{k,l},\phi_{k,l}\right),
\end{equation}
where $c_{uq}^{(n)}$ is the harmonic coefficient, $Y_u^{q}$ denotes the spherical harmonics, given by
\begin{equation}
Y_u^q(\theta,\phi) = N_u^{q}
P_u^{q}(\cos\theta)\; e^{j {q} \phi},
\; u \ge 0,\; -u \le q \le u .
\end{equation}
where $N_u^{q} = \sqrt{\frac{2u + 1}{4 \pi} \frac{(u-|q|)!}{(u+|q|)!}}$ is the normalization factor, $!$ stands for the factorial operation, $P_u^{q}(\cos\theta)$ represents the associated Legendre function of the $u$th degree and $q$th order. To make the analysis tractable, we follow the truncation operation in \cite{RA1} with a window length $T = U^2 + 2U + 1$. Therefore, (\ref{NoAppG}) can be approximated as\footnote{ Modeling and optimizing the radiation pattern via a truncated spherical-harmonic basis is an idealized model and may yield patterns that are not physically realizable. As future work, we will develop optimization methods based on the real radiation patterns of \ac{ra}s and seek orthogonal bases that are physically implementable.}
\begin{equation}
    G^{(n)}(\theta,\phi)\approx\sum_{u=0}^{U}\sum_{q=-u}^{u}c_{uq}^{(n)}Y_{u}^{q}(\theta,\phi)=\sum_{t=1}^{T}\tilde{c}_{t}^{(n)}\tilde{Y}_{t}(\theta,\phi),
\end{equation}
where $\tilde{c}^{(n)}_t = c_{uq}^{(n)}$ and $\tilde{Y}_t\left(\theta,\phi\right) = Y_u^q \left(\theta,\phi\right)$ with $t = u^2 + u + q + 1, \, u \in [0,U], \, q \in [-u,u]$. In the rest of this paper, we only consider the truncation signal model under the notation of $\mathbf{c}^{(n)} \triangleq [\tilde{c}_{1}^{(n)}, \tilde{c}_{2}^{(n)}, \ldots, \tilde{c}_{T}^{(n)}]^{\rm T} \in \mathbb{C}^{T}$, $\mathbf{b}(\theta, \phi) \triangleq [\tilde{Y}_{1}(\theta, \phi),  \ldots, \tilde{Y}_{T}(\theta, \phi)]^{\rm T} \in \mathbb{C}^{T}$. According to the energy conservation law in \cite{RA1}, an energy constraint concerning $\mathbf{c}^{(n)}$ is introduced as 
\begin{equation}
    \|\mathbf{c}^{(n)}\|_2^2 = 1 , \forall n.
\end{equation}
With the above truncated spherical harmonics decomposition, the antenna gain can be expressed as
\begin{equation}
    G^{(n)} \left(\theta,\phi\right) \approx \mathbf{b}\left(\theta,\phi\right)^{\rm H} \mathbf{c}^{(n)}.
\end{equation}
Based on the aforementioned models, the communication channel to the $k$th \ac{cu} can be formulated in a compact format as
\begin{equation}\label{comchan}
    \mathbf{h}_k  \triangleq \mathbf{F}_{\rm EM}^{\rm H} \mathbf{h}_{k}^{\rm EM} = \sqrt{\frac{N_{\rm T}}{L_k}} \sum_{l=1}^{L_k} \mathbf{F}_{\rm EM}^{\rm H} \mathbf{h}_{k,l}^{\rm EM},
\end{equation}
where
\begin{equation}
    \mathbf{F}_{\rm EM} = {\rm blkdiag} \left\{ \mathbf{c}^{(1)}, \, \mathbf{c}^{(2)}, \, \cdots, \, \mathbf{c}^{(N_{\rm T})} \right\},
\end{equation}
\begin{equation}
    \mathbf{h}_{k,l}^{\rm EM} =  \mathbf{b}\left(\theta_{k,l},\phi_{k,l}\right) \otimes \mathbf{1}_{N_{\rm T}}  \odot \left({\alpha}_{k,l} \cdot \mathbf{a}^{N_{\rm T}}_{k,l} \otimes \mathbf{1}_{T}\right).
\end{equation}

\subsubsection{Sensing Channel Model}
Since we only consider a point-like target with dominant \ac{los} path propagation in this work, the sensing channel can be modeled in the same way as the communication channel, expressed as
\begin{equation}
    \mathbf{H}_{\rm t} \triangleq \mathbf{F}_{\rm EM}^{\rm H} \mathbf{H}_{\rm t}^{\rm EM} = \mathbf{h}_{\rm t} {\tilde{\mathbf{h}}}^{\rm H}_{\rm t} = \mathbf{F}_{\rm EM}^{\rm H} \mathbf{h}^{\rm EM}_{\rm t} \tilde{\mathbf{h}}^{\rm H}_{\rm t},
\end{equation}
where $ \tilde{\mathbf{h}}_{\rm t} = \mathbf{a}^{N_{\rm R}}_{\rm t}$, $\mathbf{h}_{\rm t} = \mathbf{F}_{\rm EM}^{\rm H} \mathbf{h}_{\rm t}^{\rm EM}$, and
\begin{equation}
    \mathbf{h}_{\rm t}^{\rm EM} = \mathbf{b}\left(\theta_{\rm t},\phi_{\rm t}\right) \otimes \mathbf{1}_{N_{\rm T}}  \odot \left({\alpha}_{\rm t} \cdot \mathbf{a}_{\rm t}^{N_{\rm T}} \otimes \mathbf{1}_{T}\right),
\end{equation}
with ${\alpha}_t$ being the reflection coefficient modeled by the round-trip path-loss and radar cross section as in \cite{RCS}. The interference channels $\mathbf{H}_{{\rm int}, i}, i \in 1\dots M$, including the self-interference channel and the round-trip channel through scatterers, can be calculated similarly.

\subsection{Transmission Model}
Compared to the conventional \ac{hbf} framework, we have \ac{em} precoder $\mathbf{F}_{\rm EM}\in\mathbb{C}^{N_{\rm T}T \times N_{\rm T}}$ in addition to the baseband digital precoder $\mathbf{F}_{\rm BB} \in 
\mathbb{C}^{N_{\rm TRF} \times K}$ and analog precoder $\mathbf{F}_{\rm RF} \in 
\mathbb{C}^{N_{\rm T} \times N_{\rm TRF} }$. Thus, the communication transmission model for \ac{cu}s can be formulated as
\begin{equation}
    y_{{\rm c},k} = \mathbf{h}_k^{\rm H} \mathbf{F}_{\rm RF} [\mathbf{F}_{\rm BB}]_{:,k} \mathbf{s} + \sum\limits_{j \neq k} \mathbf{h}_k^{\rm H} \mathbf{F}_{\rm RF} [\mathbf{F}_{\rm BB}]_{:,j}  \mathbf{s} + n_k,
\end{equation}
where $\mathbf{s}$ denotes the transmitting communication signal vector, satisfying $\mathbb{E}[\mathbf{s}\mathbf{s}^{\rm H}] = \mathbf{I}_K$, $n_k$ obeys $\mathcal{CN}\left(0,\sigma_{\rm n}^2\right)$ is the \ac{awgn} at the $k$th \ac{cu}. Accordingly, the \ac{sinr} $\gamma_k$ and sum rate $R_{\rm c}$ are given by
\begin{equation}
{\gamma _k} = \frac{A_k}{B_k} =\frac{{{{\left| {{\mathbf{h}}_k^{{\text{EM}}\;{\text{H}}}{{\mathbf{F}}_{{\text{EM}}}}{\mathbf{f}_{{\rm FD},k}}} \right|}^2}}}{{\sum\limits_{j \ne k} {{{\left| {{\mathbf{h}}_k^{{\text{EM}}\;{\text{H}}}{{\mathbf{F}}_{{\text{EM}}}}{\mathbf{f}_{{\rm FD},j}}} \right|}^2} + \sigma _{\text{n}}^2} }},
\end{equation}
\begin{equation}
{R_{{\rm c}}} = \sum\limits_{k = 1}^K R_k =\sum\limits_{k = 1}^K {{{\log }_2}\left( {1 + {\gamma_{{{{k}}}}}} \right)},
\end{equation}
where ${\mathbf{f}_{{\rm FD},k}} \triangleq \mathbf{F}_{\rm RF}{{[{{\mathbf{F}}_{{\text{BB}}}}]}_{:,k}} $. For the radar sensing transmission model, the reflected echo signal at the radar receiver is expressed as
\begin{equation}\label{RTE}
{\mathbf{y}_{\rm r}} = \underbrace {{\mathbf{H}_{\rm t}^{\rm H} \mathbf{x}}}_{{\text{Target signal}}} + \underbrace {\sum\limits_{m = 1}^M {{{\mathbf{H}}^{\rm H}_{{\text{int,}}m}}{\mathbf{x}}} }_{{\text{Scatterer Interference}}} + {\mathbf{n}},
\end{equation}
where $\mathbf{x} = \mathbf{F}_{\rm RF}\mathbf{F}_{\rm BB}\mathbf{s}$, and $\mathbf{n}$ denotes the \ac{awgn} vector obeying $\mathcal{CN}(\mathbf{0},\sigma_{\rm n}^2\mathbf{I}_{N_{\rm R}})$. Accordingly, the radar \ac{scnr} $\eta$ can be formulated as
\begin{equation}\label{scnrequ}
\eta  = \sum\limits_{k = 1}^K {\frac{{{C_k}}}{D}}  = \sum\limits_{k = 1}^K {\frac{{\left| {{\mathbf{h}}_{\text{t}}^{{\text{EM}}\;{\text{H}}}{{\mathbf{F}}_{{\text{EM}}}}{{\mathbf{f}}_{{\text{FD}},k}}} \right|_2^2}}{{\sum\limits_{j = 1}^K {\sum\limits_{m = 1}^M {| {{\mathbf{h}}_{\operatorname{int} ,m}^{{\text{EM}}\;{\text{H}}}{{\mathbf{F}}_{{\text{EM}}}}{{\mathbf{f}}_{{\text{FD}},j}}} |_2^2} }  + \sigma _{\text{n}}^2}}}.
\end{equation}
\subsection{Problem Formulation}
With the given transmitting power budget, antenna radiation energy, and phase shift modulus-1 constraints, the baseband digital precoder $\mathbf{F}_{\rm BB}$, phase shift analog precoder $\mathbf{F}_{\rm RF}$, and \ac{era} \ac{em} precoder $\mathbf{F}_{\rm EM}$ are optimized to maximize the weighted sum of the communication sum rate and the \ac{scnr}. Mathematically, the optimization problem is cast as follows
\begin{subequations}\label{optp1}
\begin{align}
&\mathcal{P}_1:\;\mathop {\max }\limits_{{\mathbf{F}_{\rm EM}},{\mathbf{F}_{\rm RF}},{\mathbf{F}_{\rm BB}}} \tilde{\beta} R_{\rm c} + \beta \eta\\
&{\rm s.t.} \ \ {\rm Tr}\left({{\mathbf{F}}_{{\text{RF}}}}{{\mathbf{F}}_{{\text{BB}}}}{\mathbf{F}}_{{\text{BB}}}^{\text{H}}{\mathbf{F}}_{{\text{RF}}}^{\text{H}}\right) \leq P, \\
&\ \ \ \ \     \|\mathbf{c}^{(n)}\|_2^2 = 1 , \forall n,\\
&\ \ \ \ \     |[\mathbf{F}_{\rm RF}]_{i,j}| = 1, \forall i,j,
\end{align}%
\end{subequations}
where $\tilde{\beta}$ and $\beta$ are trade-off factors between communication and sensing, satisfying $\tilde{\beta} + \beta = 1$. Due to the sum of fractions in the communication sum rate, constant modulus, and norm constraints, this problem is highly nonconvex.

\section{Proposed Tri-hybrid Beamforming Method}

\begin{figure*}[hb]
\small
\centering
\rule{\linewidth}{1pt} 
\begin{equation}\label{BFOptSol}
{\mathbf{f}}_{{\text{FD,}}k}^ \star  = {\left[ {\sum\limits_{j = 1}^K {{{\left| {{p_j}} \right|}^2}{{\mathbf{h}}_j}{\mathbf{h}}_j^{\text{H}}}  + \left( {\sum\limits_{j = 1}^K {{{\left| {{q_j}} \right|}^2}} } \right)\sum\limits_{m = 1}^M {{\mathbf{h}}_{\operatorname{int} ,m}^{}{\mathbf{h}}_{\operatorname{int} ,m}^{\text{H}} + {\mu ^ \star }{\mathbf{I}}} } \right]^\dag }\left( {{p_k}\sqrt {\tilde \beta \left( {1 + {\gamma _k}} \right)} {{\mathbf{h}}_k} + {q_k}\sqrt \beta  {{\mathbf{h}}_{\text{t}}}} \right),\tag{27}
\end{equation}

\begin{equation}\label{Euclgrad}
\begin{aligned}
    {\nabla _{{{ {\mathbf{c}}}^{(n)}}}}f_{{\text{qua}}}^{(n)} &= \sum\limits_{k = 1}^K { \left[ {\sum\limits_{j = 1}^K {{{\left| {{p_k}} \right|}^2}{\mathbf{\tilde h}}_{k,j,n}^{{\text{EM}}}{{\left( {{\mathbf{\tilde h}}_{k,j,n}^{{\text{EM}}}} \right)}^{\text{H}}} + \sum\limits_{m = 1}^M {{{\left| {{q_k}} \right|}^2}{\mathbf{\tilde h}}_{\operatorname{int} ,j,m,n}^{{\text{EM}}}{{\left( {{\mathbf{\tilde h}}_{\operatorname{int} ,j,m,n}^{{\text{EM}}}} \right)}^{\text{H}}}} } } \right]} {{{\mathbf{ c}}}^{(n)}} - \\
   & \left[ {{p_k}\sqrt {\tilde \beta \left( {1 + {\gamma _k}} \right)} {\mathbf{\tilde h}}_{k,k,n}^{{\text{EM}}} + {q_k}\sqrt \beta   - \sum\limits_{j = 1}^K {\left( {{{\left| {{p_k}} \right|}^2}{\text{con}}{{\text{s}}_{k,j,n}}{\mathbf{\tilde h}}_{k,j,n}^{{\text{EM}}} - \sum\limits_{m = 1}^M {{{\left| {{q_k}} \right|}^2}{{\mathop {{\text{cons}}}\limits^ \leftrightarrow  }_{j,m,n}}{\mathbf{\tilde h}}_{\operatorname{int} ,j,m,n}^{{\text{EM}}}} } \right)} } \right].
\end{aligned} \tag{35}
\end{equation}
\end{figure*}

Under the \ac{ao} framework, problem (\ref{optp1}) is transformed through the \ac{fp} technique, then the \ac{mo} will be employed to tackle the constant modulus/norm constraints.

\subsection{Optimizing $\mathbf{F}_{\rm RF}$ and $\mathbf{F}_{\rm BB}$ with Given $\mathbf{F}_{\rm EM}$}
With given $\mathbf{F}_{\rm EM}$, problem (\ref{optp1}) becomes a classical ISAC beamforming problem with main challenges lying in the multiple sums in the objective function. Using the Lagrangian dual transform in \cite{fp} with respect to the communication sum rate part of the objective function, we have a new objective as
\begin{equation}\label{LagObj}
\small
{f_{{\text{Lag}}}} = \sum\limits_{k = 1}^K { \tilde{\beta} \left[ {\log \left( {1 + {\gamma _k}} \right) - {\gamma _k} + \frac{{\left( {1 + {\gamma _k}} \right){A_k}}}{{{A_k} + {B_k}}}} \right]}  + \beta \frac{{{C_k}}}{D}.
\end{equation}
Furthermore, the quadratic transform can be applied to (\ref{LagObj}), yielding
% \begin{equation}\label{QuaObj}
% \begin{aligned}
% &f_{\rm qua} = \sum\limits_{k = 1}^K \log \left( {1 + {\gamma _k}} \right) - {\gamma _k} + 2\Re \left( {p_k^*\sqrt {\left( {1 + {\gamma _k}} \right){A_k}} } \right)  \\
% &- {{\left| {{p_k}} \right|}^2}\left( {{A_k} + {B_k}} \right)  + 2\beta \Re \left( {q_k^*\sqrt C_k } \right) -  {\left| {{q_k}} \right|^2}\beta (C_k+D).
% \end{aligned}
% \end{equation}
{\small
\begin{flalign}\label{QuaObj}
&f_{\rm qua} = \sum\limits_{k = 1}^K  \tilde{\beta} \log \left( 1 + {\gamma _k} \right) - \tilde{\beta} {\gamma _k} 
+ 2\Re \left( p_k^*\sqrt { \tilde{\beta} (1 + {\gamma _k}) {A_k} } \right) - \notag \\
& \left| p_k \right|^2 \left( A_k + B_k \right) 
+ 2 \Re \left( q_k^*\sqrt{\beta C_k} \right) 
- \left| q_k \right|^2 D.
\end{flalign}
}

Owing to the multi-fraction decomposition of \ac{scnr} in (\ref{scnrequ}), the square root of $C_k$ can be smoothly written as a linear term in $\mathbf{f}_{{\rm FD},k}$. Now, with the new objective function in (\ref{QuaObj}), the new optimization is cast as follows
\begin{subequations}\label{optp2}
\begin{align}
&\mathcal{P}_2:\;\mathop {\max }\limits_{{\mathbf{f}_{{\rm FD},k}},\gamma_k,p_k,q_k}\; f_{\rm qua} \left(\mathbf{f}_{{\rm FD},k},\gamma_k,p_k,q_k\right)\\
&{\rm s.t.} \ \ \sum\limits_{k = 1}^K {{\text{Tr}}\left( {{{\mathbf{f}}_{{\text{FD},k}}}{\mathbf{f}}_{{\text{FD},k}}^{\text{H}}} \right) \leqslant P}.
\end{align}%
\end{subequations}
Similar to the conventional \ac{fp} problem, we have closed-form optimal solutions for $\gamma_k,p_k,q$ as follows
\begin{equation}\label{gammaupdate1}
\gamma _k^ \star  = \frac{{{{\left| {{\mathbf{h}}_k^{\text{H}}{{\mathbf{f}}_{{\text{FD,}}k}}} \right|}^2}}}{{\sum\limits_{j \ne k} {{{\left| {{\mathbf{h}}_k^{\text{H}}{{\mathbf{f}}_{{\text{FD,}}j}}} \right|}^2} + \sigma _{\text{n}}^2} }},
\end{equation}
\begin{equation}\label{pupdate1}
p_k^ \star  = \frac{{\sqrt {\tilde \beta \left( {1 + {\gamma _k}} \right)} {\mathbf{h}}_k^{\text{H}}{{\mathbf{f}}_{{\text{FD,}}k}}}}{{\sum\limits_{j = 1}^K {{{\left| {{\mathbf{h}}_k^{\text{H}}{{\mathbf{f}}_{{\text{FD,}}j}}} \right|}^2} + \sigma _{\text{n}}^2} }},
\end{equation}
\begin{equation}\label{qupdate1}
q_k^ \star  = \frac{ \sqrt{\beta}{{\mathbf{h}}_{\text{t}}^{\text{H}}{{\mathbf{f}}_{{\text{FD,}}k}}}}{{\sum\limits_{j = 1}^K {\sum\limits_{m = 1}^M {{{\left| {{\mathbf{h}}_{\operatorname{int} ,m}^{\text{H}}{{\mathbf{f}}_{{\text{FD,}}j}}} \right|}^2}} }  + \sigma _{\text{n}}^2}}.
\end{equation}
The closed form of the optimal beamformer $\mathbf{f}_{{\rm FD},k}^{\star}$ is given in (\ref{BFOptSol}), where $\mu^{\star}$ is the optimal Lagrangian dual variable of the power budget constraint obtained by the line search.
% Since the \ac{scnr} in (\ref{scnrequ}) involves all $\mathbf{f}_{{\rm FD},k}, \; \forall k$, it is hard to give the closed-form solution of $\mathbf{f}_{{\rm FD},k}$ like the conventional \ac{mu} \ac{mimo} beamforming problem. Instead, the conjugate gradient is given in (\ref{conjgrad}) for the projected gradient descent. 
Furthermore, the analog beamformer $\mathbf{F}_{\rm RF}$ and digital beamformer $\mathbf{F}_{\rm BB}$ are obtained by the following optimization problem
\setcounter{equation}{27}
\begin{subequations}\label{MO}
\begin{align}
\mathcal{P}_3:\;&\mathop {\min }\limits_{\mathbf{F}_{\rm RF},\;\mathbf{F}_{\rm BB}} \|\mathbf{F}_{\rm RF}\mathbf{F}_{\rm BB} - \mathbf{F}_{\rm FD}\|_F^2,\\
&{\rm{s}}{\rm{.t}}{\rm{.}}\; \left|\left[\mathbf{F}_{\rm RF}\right]_{i,j}\right| = 1,\; \forall i,\;j,\\
&\;{\kern 1pt} \;{\kern 1pt} \;{\kern 1pt} \;  \; \|\mathbf{F}_{\rm RF}\mathbf{F}_{\rm BB} \|_F^2  \leq P.
\end{align}
\end{subequations}
This is a classic approximation in the hybrid beamforming field, which can be solved by the \ac{mo} method given in \cite{mo}. Thus, the details are omitted here for simplicity.
\subsection{Optimizing $\mathbf{F}_{\rm EM}$ with Given $\mathbf{F}_{\rm RF}$ and $\mathbf{F}_{\rm BB}$}
Given $\mathbf{F}_{\rm RF}$ and $\mathbf{F}_{\rm BB}$, we conduct an \ac{ao} among different antennas. Focusing on the $n$th antenna, we have the following equivalent expression
\begin{flalign}\label{equivalentEqu}
&{({\mathbf{h}}_k^{{\text{EM}}})^{\text{H}}}{\mathbf{F}}_{{\text{EM}}}^{} \mathbf{f}_{{\rm FD},k} = \sum\limits_{n = 1}^{{N_t}} {{f_{{\text{FD}},k,(n)}}{{({\mathbf{h}}_{k,(n)}^{{\text{EM}}})}^{\text{H}}}{{\mathbf{c}}^{(n)}}} \notag \\
&= {f_{{\text{FD}},k,(n)}}{({\mathbf{h}}_{k,(n)}^{{\text{EM}}})^{\text{H}}}{{\mathbf{c}}^{(n)}} + \sum\limits_{m \ne n} {{f_{{\text{FD}},k,(m)}}{{({\mathbf{h}}_{k,(m)}^{{\text{EM}}})}^{\text{H}}}{{\mathbf{c}}^{(m)}}} \notag \\
&\triangleq {({\mathbf{\tilde h}}_{k,k,n}^{{\text{EM}}})^{\text{H}}}{{\mathbf{c}}^{(n)}} + {\text{con}}{{\text{s}}_{k,k,n}}, \hfill 
\end{flalign}
with ${\mathbf{h}}_k^{{\text{EM}}} = {[ {\begin{array}{*{20}{c}}
  {{\mathbf{h}}_{k,(1)}^{{\text{EM}}\;{\text{T}}}}&{{\mathbf{h}}_{k,(2)}^{{\text{EM}}\;{\text{T}}}}& \cdots &{{\mathbf{h}}_{k,({N_{\text{T}}})}^{{\text{EM}}\;{\text{T}}}} 
\end{array}} ]^{\text{T}}}$ and ${f_{{\text{FD}},k,(n)}}$ being the $n$th element of $\mathbf{f}_{{\rm FD},k}$. With (\ref{equivalentEqu}) at hand, the \ac{sinr} and the \ac{scnr} can be formulated as
\begin{equation}\label{newsinr}
{\gamma _k^{(n)}} = \frac{A_k^{(n)}}{B^{(n)}_k} = \frac{{|{{({\mathbf{\tilde h}}_{k,k,n}^{{\text{EM}}})}^{\text{H}}}{{\mathbf{c}}^{(n)}} + {\text{con}}{{\text{s}}_{k,k,n}}{|^2}}}{{\sum\limits_{j \ne k} | {{({\mathbf{\tilde h}}_{k,j,n}^{{\text{EM}}})}^{\text{H}}}{{\mathbf{c}}^{(n)}} + {\text{con}}{{\text{s}}_{k,j,n}}{|^2} + \sigma _n^2}},
\end{equation}
\begin{flalign}\label{newscnr}
&{\eta ^{(n)}} = \sum\limits_{k = 1}^K {\frac{{C_k^{(n)}}}{{{D^{(n)}}}}}  \notag\\
&= \sum\limits_{k = 1}^K {\frac{{|{{({\mathbf{\tilde h}}_{{\text{t}},k,n}^{{\text{EM}}})}^{\text{H}}}{{\mathbf{c}}^{(n)}} + {{\mathop {{\text{cons}}}\limits^ \leftrightarrow  }_{k,n}}{|^2}}}{{\sum\limits_{j = 1}^K {\sum\limits_{m = 1}^M {|{{({\mathbf{\tilde h}}_{{\text{int}},j,m,n}^{{\text{EM}}})}^{\text{H}}}{{\mathbf{c}}^{(n)}} + {{\mathop {{\text{cons}}}\limits^ \leftrightarrow  }_{j,m,n}}{|^2}} }  + \sigma _{\text{n}}^2}}} ,
\end{flalign}
where ${\text{con}}{{\text{s}}_{k,j,n}} = \sum\limits_{m \ne n} {{{({\mathbf{h}}_{k,(m)}^{{\text{EM}}})}^{\text{H}}}{{\mathbf{c}}^{(m)}}{f_{{\text{FD}},j,(m)}}}$, ${\mathbf{\tilde h}}_{k,j,n}^{{\text{EM}}} = {f_{{\text{FD}},j,(n)}}^*{\mathbf{h}}_{k,(n)}^{{\text{EM}}}$, ${\mathbf{\tilde h}}_{{\text{t}},k,n}^{{\text{EM}}} = {f_{{\text{FD}},k,(n)}^*}{\mathbf{h}}_{{\text{t}},(n)}^{{\text{EM}}}$, ${\mathbf{\tilde h}}_{{\text{int}},j,m,n}^{{\text{EM}}} = {f_{{\text{FD}},j,(n)}^{*}}{\mathbf{h}}_{{\text{int}},m,\left( n \right)}^{{\text{EM}}}$, ${\mathop {{\text{cons}}}\limits^ \leftrightarrow  _{k,n}} = \sum\limits_{m \ne n} {{f_{{\text{FD}},k,(m)}}{{({\mathbf{h}}_{{\text{t}},(m)}^{{\text{EM}}})}^{\text{H}}}{{\mathbf{c}}^{(m)}}}$, and ${\mathop {{\text{cons}}}\limits^ \leftrightarrow  _{j,m,n}} = \sum\limits_{i \ne n} {{f_{{\text{FD}},j,(i)}}{{({\mathbf{h}}_{{\text{int}},m,\left( i \right)}^{{\text{EM}}})}^{\text{H}}}{{\mathbf{c}}^{(i)}}} $. Thus, the quadratic transform objective function (\ref{QuaObj}) can be rewritten with respect to $\mathbf{c}^{\left(n\right)}$ with the new numerators $A_k^{(n)}$, $C_k^{(n)}$ as well as denominators $B_k^{(n}$, $D^{(n)}$.
%shown in (\ref{newsinr}) and (\ref{newscnr}).
% \begin{equation}\label{QuaObj}
% \begin{aligned}
% f_{{\text{qua}}}^{(n)} = \sum\limits_{k = 1}^K {\log } \left( {1 + \gamma _k^{(n)}} \right) - \gamma _k^{(n)} + 2\Re \left( {p_k^{(n)\;*}\sqrt {\left( {1 + \gamma _k^{(n)}} \right)A_k^{(n)}} } \right) \hfill \\
% - {\left| {{p_k}} \right|^2}\left( {A_k^{(n)} + B_k^{(n)}} \right) + 2\beta \Re \left( {{q^{(n)\;*}}\sqrt {{C^{(n)}}} } \right) - {\left| {{q^{(n)}}} \right|^2}\beta {D^{(n)}}. \hfill 
% \end{aligned}
% \end{equation}
Similarly, the optimal solutions for the optimized variables are given by
\begin{equation}\label{gammaupdate2}
\gamma _k^{(n)\; \star } = \frac{{|{{({\mathbf{\tilde h}}_{k,k,n}^{{\text{EM}}})}^{\text{H}}}{{\mathbf{c}}^{(n)}} + {\text{con}}{{\text{s}}_{k,k,n}}{|^2}}}{{\sum\limits_{j \ne k} | {{({\mathbf{\tilde h}}_{k,j,n}^{{\text{EM}}})}^{\text{H}}}{{\mathbf{c}}^{(n)}} + {\text{con}}{{\text{s}}_{k,j,n}}{|^2} + \sigma _n^2}},
\end{equation}

\begin{equation}\label{pupdate2}
p_k^{(n)\; \star } = \frac{{\sqrt  { \tilde{\beta} \left( {1 + \gamma _k^{(n)}} \right)} \left[ {{({\mathbf{\tilde h}}_{k,k,n}^{{\text{EM}}})}^{\text{H}}}{{\mathbf{c}}^{(n)}} + {\text{con}}{{\text{s}}_{k,k,n}} \right]}}{{\sum\limits_{j =1 }^K | {{({\mathbf{\tilde h}}_{k,j,n}^{{\text{EM}}})}^{\text{H}}}{{\mathbf{c}}^{(n)}} + {\text{con}}{{\text{s}}_{k,j,n}}{|^2} + \sigma _n^2}},
\end{equation}
\begin{equation}\label{qupdate2}
q_k^{(n)\; \star } = \frac{{\sqrt \beta  \left[{{({\mathbf{\tilde h}}_{{\text{t}},k,n}^{{\text{EM}}})}^{\text{H}}}{{\mathbf{c}}^{(n)}} + {{\mathop {{\text{cons}}}\limits^ \leftrightarrow  }_{k,n}}\right]}}{{\sum\limits_{j = 1}^K {\sum\limits_{m = 1}^M {|{{({\mathbf{\tilde h}}_{{\text{int}},j,m,n}^{{\text{EM}}})}^{\text{H}}}{{\mathbf{c}}^{(n)}} + {{\mathop {{\text{cons}}}\limits^ \leftrightarrow  }_{j,m,n}}{|^2}} }  + \sigma _{\text{n}}^2}}.
\end{equation}

Furthermore, the optimal solution of $\mathbf{c}^{(n)}$ is obtained through the \ac{mo} since constraint (\ref{optp1}c) is actually a sphere-manifold constraint. We first give the Euclidean gradient as shown in (\ref{Euclgrad}), then the Riemannian gradient can be obtained through the following projection operation
\setcounter{equation}{35}
\begin{equation}\label{Riegrad}
{\rm grad} f_{\rm qua}^{{\mathbf{c}}^{(n)}} = {\nabla _{{{\mathbf{c}}^{(n)}}}}f_{{\text{qua}}}^{(n)} - \left({{{\mathbf{c}}^{(n)\; {\rm T}}}} {\nabla _{{{\mathbf{c}}^{(n)}}}}f_{{\text{qua}}}^{(n)} \right){{{\mathbf{c}}^{(n)}}},
\end{equation}
where ${{\tilde{\mathbf{c}}^{(n)}}} = {{{\mathbf{c}}^{(n)}}}/\sqrt{4\pi}$. Using the Riemannian gradient and step size $\delta$, the optimized variable can be updated as follows
\begin{equation}\label{cnupdate}
{\mathbf{ c}}^{(n)} = \frac{{{\mathbf{ c}}^{(n)} - \delta {\text{grad}}f_{{\text{qua}}}^{ {\mathbf{c}}^{(n)}}}}{{\left\| {{\mathbf{ c}}^{(n)} - \delta {\text{grad}}f_{{\text{qua}}}^{ {\mathbf{c}}^{(n)}}} \right\|}},
\end{equation}
where the step size $\delta$ is determined by the conventional Armijo rule. The overall algorithm for addressing problem (\ref{optp1}) is outlined in Algorithm \ref{alg1}. The main complexity of this algorithm comes from the inverse matrix computation in (\ref{BFOptSol}) and Euclidean gradient computation in (\ref{Euclgrad}), the orders of which are $\mathcal{O}(N_{\rm T}^3)$ and $\mathcal{O}(KT+MT)$, respectively. Thus, with the given \ac{ao} number of iterations $I_{\rm AO}$, the overall computational complexity order is $\mathcal{O}[I_{\rm AO}(KN_{\rm T}^3 + KN_{\rm T}(KT+MT))]$.

\begin{algorithm}[]
	\color{black}
\caption{Proposed tri-hybrid beamforming optimization scheme for Problem (\ref{optp1}).}\label{alg1}
\begin{algorithmic}[1]
		\STATE {\bf Initialize:} Initial feasible tri-hybrid beamformer: $\mathbf{F}_{\rm RF}^0$, $\mathbf{F}_{\rm BB}^0$, $\mathbf{F}_{\rm EM}^0$.
		\REPEAT
		 \STATE Update $\gamma_k, \, p_k, \, q_k,\,k = 1,\dots,K$ according to (\ref{gammaupdate1})-(\ref{qupdate1}),
      % \STATE Update $p_k,\,k = 1,\dots,K$ according to (\ref{pupdate1}),
      % \STATE Update $q_k,\,k = 1,\dots,K$ according to (\ref{qupdate1}),
      \STATE Update the fully digital beamformer $\mathbf{f}_{{\rm FD},k},\,k = 1,\dots,K$ according to (\ref{BFOptSol}),
      \STATE Update the hybrid beamformer $\mathbf{F}_{{\rm RF}}$ and $\mathbf{F}_{{\rm BB}}$ according to problem (\ref{MO}),
      \FOR{$n=1$ to $N_{\rm T}$}
		 \STATE Update $\gamma^{(n)}_k,\, p^{(n)}_k,\, q^{(n)}_k,\, k = 1,\dots,K$ according to (\ref{gammaupdate2}), (\ref{pupdate2}), and (\ref{qupdate2}),
      % \STATE Update $p^{(n)}_k,\,k = 1,\dots,K$ according to (\ref{pupdate2}),
      % \STATE Update $q^{(n)}_k,\,k = 1,\dots,K$ according to (\ref{qupdate2}),
      \STATE Update the \ac{em} beamformer $\mathbf{c}^{(n)}$ according to (\ref{cnupdate}),
      \ENDFOR
	   \UNTIL the objective value converges or reaches the maximum iteration number.
\end{algorithmic}
\end{algorithm}

\section{Simulation Results}
In this section, simulation results are presented to validate the advantages of the proposed ERA-ISAC scheme in terms of achievable sum rate and \ac{scnr} performance. The specific simulation parameters are provided in Table \ref{SimSetup}. The \ac{cu}, targets, and scatterers were independently and uniformly distributed in spherical coordinates, with elevation angle $\theta \in (0^{\rm o},180^{\rm o})$, azimuth angle $\phi \in (0^{\rm o},360^{\rm o})$, and range $r \in (10 \,{\rm m},50 \,{\rm m})$. Based on these positions, we generated 1,000 independent channel trials.

\renewcommand{\arraystretch}{1.1}
\begin{table}[h]
	\caption{Simulation parameter setup.}
	\label{DMoverview}
	\centering
	\begin{threeparttable}
		\begin{tabular}{*{3}{>{\centering\arraybackslash}m{2cm}>{\centering\arraybackslash}m{1.8cm}>{\centering\arraybackslash}m{3.8cm}}} 
			\Xhline{2pt}
			\textbf{Parameter} & \textbf{Value} & \textbf{Description} \\
			\Xhline{1pt}
			$N_{\rm T}$ &  $16 = 4 \times 4$  & Transmit antenna number  \\
			$N_{\rm R}$ & $16  = 4\times 4$ &   Receive antenna number\\ 
            $N_{\rm TRF}$ &  2               & Transmit RF chain number \\
			$f_c$ & 3 GHz & Carrier frequency \\
			$d_x = d_y = \lambda/2$ & 0.05 m  &  Antenna element spacing\\
            $K$ & 2 & \ac{cu} number \\
			$M $ & 2 & Scatterer number \\
            $T$ & 25 & Truncation length \\
            $\sigma_{\rm n}^2$ & -80 dBm & Noise power\\
			\Xhline{2pt}
		\end{tabular}%
	\end{threeparttable}
	\label{SimSetup}%
\end{table}%

\begin{figure}[ht]
	\centering
	\begin{minipage}{0.24\textwidth}
		\centering
        \includegraphics[scale=0.7]{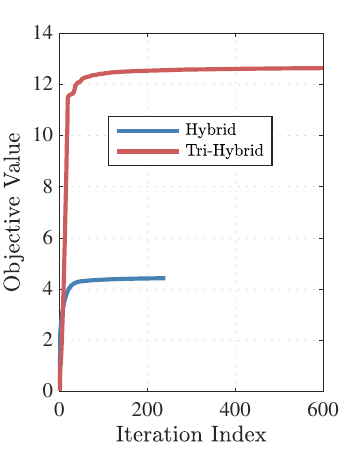}
		\subcaption*{(a)}
		\vspace{-1mm}
	\end{minipage}
	\hfill
	\begin{minipage}{0.24\textwidth}
		\centering
        \includegraphics[scale=0.7]{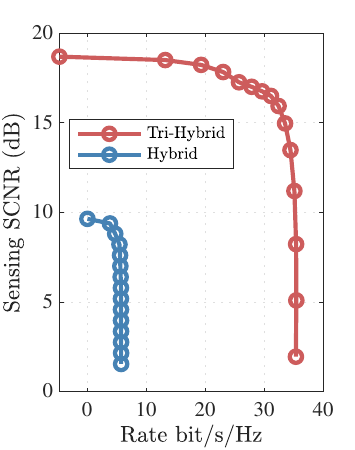}
		\subcaption*{(b)}
		\vspace{-1mm}
	\end{minipage}
	\caption{(a) The weighted sum of sum rate and sensing SCNR versus iteration number; (b) \ac{sandc} trade-off comparison between \ac{era}-\ac{isac} and \ac{oa}-\ac{isac} systems when the transmit power equals -20 dBm.}\label{simfig1}
	\vspace{-3mm}
\end{figure}

\begin{figure}[ht]
	\centering
	\includegraphics[scale=0.7]{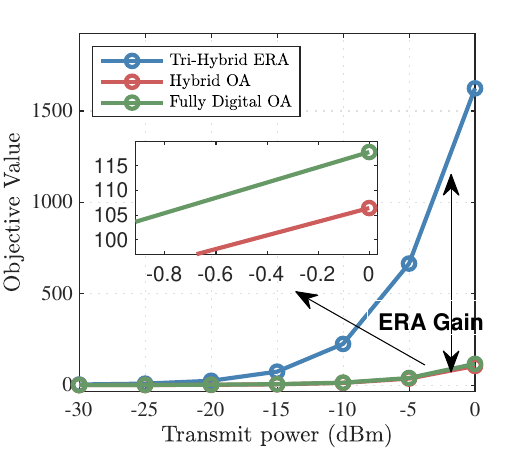}
	\caption{The weighted sum of sum rate and sensing SCNR versus transmit power density when $\beta = 0.5$.}
	\label{simfig2}
\end{figure}

\begin{figure}[ht]
	\centering
	\begin{minipage}{0.24\textwidth}
		\centering
        \includegraphics[scale=0.8]{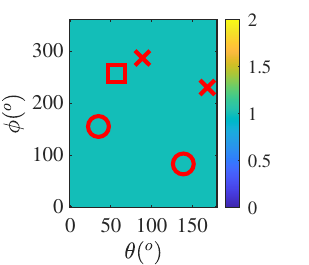}
		\subcaption*{(a) single OA}
	\end{minipage}
	\hfill
	\begin{minipage}{0.24\textwidth}
		\centering
        \includegraphics[scale=0.8]{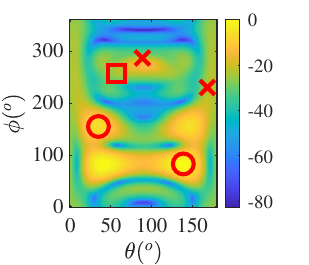}
		\subcaption*{(b) OA array (dB)}
	\end{minipage}
	\hfill
	\begin{minipage}{0.24\textwidth}
		\centering
        \includegraphics[scale=0.8]{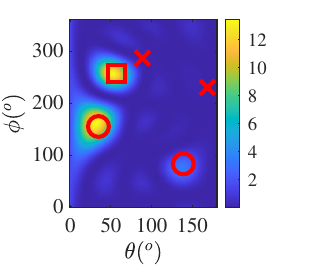}
		\subcaption*{(c) single ERA}
	\end{minipage}
	\hfill
	\begin{minipage}{0.24\textwidth}
		\centering
        \includegraphics[scale=0.8]{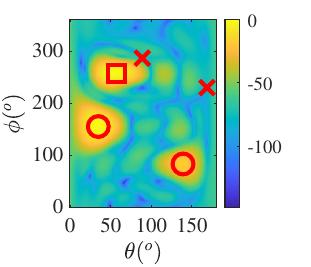}
		\subcaption*{(d) ERA array (dB)}
	\end{minipage}
	\caption{Radiation patterns when $\beta = 0.5$ and transmit power density equals -20 dBm (circle: \ac{cu}, square: target, cross: scatterers). }
	\label{simfig4}
\end{figure}

Fig. \ref{simfig1} (a) illustrates the convergence behavior of the proposed tri-hybrid beamforming scheme with \ac{era}, in comparison with its conventional hybrid beamforming counterpart with \ac{oa}. Both schemes demonstrate fast convergence, attributed to the \ac{fp}. In Fig. \ref{simfig1} (b), the trade-off between communication and sensing is presented. The proposed tri-hybrid beamforming scheme with \ac{era} strikes a significantly better trade-off in \ac{sandc} performance.

Fig. \ref{simfig2} presents the optimization objective value, i.e., the weighted sum of sum rate and sensing \ac{scnr}, versus transmit power of the tri-hybrid, hybrid, and fully digital beamforming schemes. By optimizing the radiation pattern of \ac{era}, the proposed tri-hybrid beamforming scheme achieves substantial performance gains over the conventional hybrid beamforming approach, even surpassing the fully digital scheme. This demonstrates the great potential of \ac{era} in enhancing performance for \ac{isac} systems. To provide a more intuitive illustration, Fig. \ref{simfig4} depicts the single-antenna and the array radiation patterns of both the conventional \ac{oa} and \ac{era}. It can be observed that \ac{era} exhibits superior beamforming capability in both single-antenna and array patterns.

\section{Conclusion}
This paper proposes an \ac{era}-\ac{isac} system and develops a tri-hybrid beamforming scheme. The proposed approach exhibits fast convergence, attains a superior communication–sensing trade-off, and outperforms both conventional hybrid and even fully digital counterparts, demonstrating the strong potential of \ac{era}s for future \ac{mimo}-\ac{isac} deployments. As future work, we will pursue hardware-aware optimization and investigate a realizable orthogonal basis for \ac{era}s.

\bibliography{IEEEabrv,references}

\end{document}